\documentclass[a4paper,aps,prl,floatfix,twocolumn,superscriptaddress,nobibnotes]{revtex4}

\usepackage[pdftex]{graphicx}
\setkeys{Gin}{width=0.9\columnwidth}

\usepackage{times}
\usepackage{verbatim}
\usepackage{bbold}
\usepackage{bbm}

\usepackage{amsmath}
\usepackage{amsfonts}
\usepackage{color}

\usepackage[english]{babel}
\usepackage{microtype}
\usepackage{comment}
\usepackage[normalem]{ulem}
\usepackage{float}
\usepackage{xspace}

\usepackage{relsize}

\usepackage[hidelinks,unicode=true]{hyperref}
\hypersetup{colorlinks=true,linkcolor=blue,urlcolor=blue,citecolor=blue,pdfhighlight=/N
}

\usepackage{lipsum}

\PassOptionsToPackage{hyphens}{url}

\newcommand{\R}[1]{\texttt{\small{r/#1}}}
\newcommand{\V}[1]{\texttt{\small{v/#1}}}

\newcommand{\GA}{\emph{GreatAwakening}\xspace}
\newcommand{\FPH}{\emph{FatPeopleHate}\xspace}

\newcommand{\rGA}{\R{GreatAwakening}\xspace}
\newcommand{\rFPH}{\R{FatPeopleHate}\xspace}

\newcommand{\vGA}{\V{GreatAwakening}\xspace}

\newcommand{\sGA}{\ensuremath{\operatorname{GA}}\xspace}
\newcommand{\sFPH}{\ensuremath{\operatorname{FPH}}\xspace}

\newcommand{\spara}[1]{\smallskip\noindent{\bf #1}}

\newcommand{\squishlist}{
 \begin{list}{$\bullet$}
  {  \setlength{\itemsep}{0pt}
     \setlength{\parsep}{3pt}
     \setlength{\topsep}{3pt}
     \setlength{\partopsep}{0pt}
     \setlength{\leftmargin}{2em}
     \setlength{\labelwidth}{1.5em}
     \setlength{\labelsep}{0.5em}
} }

\newcommand{\num}[1]{#1}

\begin{document}

\title[Online conspiracy communities are more resilient to deplatforming]{Online conspiracy communities are more resilient to deplatforming}

\author{Corrado Monti}
\affiliation{CENTAI Institute, Torino, Italy}

\author{Matteo Cinelli}
\affiliation{Department of Computer Science, Sapienza University of Rome, Italy}

\author{Carlo Valensise}
\affiliation{Centro Ricerche Enrico Fermi, Rome, Italy}

\author{Walter Quattrociocchi}
\thanks{Corresponding author: walter.quattrociocchi@uniroma1.it}
\affiliation{Department of Computer Science, Sapienza University of Rome, Italy}

\author{Michele Starnini}
\thanks{Corresponding author: michele.starnini@gmail.com}
\affiliation{Departament de Fisica, Universitat Politecnica de Catalunya, Campus Nord, 08034 Barcelona, Spain}
\affiliation{CENTAI Institute, Torino, Italy}

\begin{abstract}
Online social media foster the creation of active communities around shared narratives.
Such communities may turn into incubators for conspiracy theories -- some spreading violent messages that could sharpen the debate and potentially harm society.
To face these phenomena, most social {media} platforms implemented moderation policies, ranging from posting warning labels up to deplatforming, i.e., permanently banning users. %
Assessing the effectiveness of content moderation is crucial for balancing societal safety while preserving the right to free speech.
 In this paper, we compare the shift in behavior of users affected by the ban of two large communities on Reddit, \GA and \FPH, which were dedicated to spreading the QAnon conspiracy and body-shaming individuals, respectively.
 Following the ban, both communities partially migrated to Voat, an unmoderated Reddit clone. %
 We estimate how many users migrate, finding that users in the conspiracy community are much more likely to leave Reddit altogether and join Voat. %
Then, we quantify the behavioral shift within Reddit and across Reddit and Voat by matching common users.
Few migrating zealots 
drive the growth of the new \GA community on Voat, while this effect is absent for \FPH.
 Finally, conspiracy users migrating from Reddit tend to recreate their previous social network on Voat. %
 Our findings suggest that banning conspiracy communities hosting violent content should be carefully designed, as these communities may be more resilient to deplatforming.
\end{abstract}

\maketitle

\section{Introduction}

Conspiracy theories posit that significant sociopolitical events are the result of a deliberate and coordinated scheme by a small group of influential individuals~\cite{douglas2019understanding}.
While known historical conspiracies are limited in time, scope, number of actors, and complicacy, scholars distinguished a ``conspiracy mindset'' characterized by Hofstader's paranoid style: the belief in ``a vast, gigantic and yet subtle machinery of influence set in motion to undermine and destroy a way of life''~\cite{van2021paranoid,hofstadter2012paranoid}.
The QAnon conspiracy theory is a prominent example: 
its adherents believe in a sinister global cabal consisting of powerful individuals who engage in cannibalism and pedophilia, and supposedly conspired against former U.S. President Donald Trump~\cite{kline2021eat, bleakley2021panic}.
Emerged in 2017 as a grouping of far-right conspiracy theories, it developed a huge online following.
Some of its believers have since been responsible of violent acts, harming those around them~\cite{amarasingam2020qanon,jensen2021qanon}.
It has even been defined as an addiction-inducing cult, that can destroy one's life~\cite{hacker2021socio}.
Online social media, where users can easily join communities around shared narratives \cite{Sunstein_2002,del2016spreading,Bail_2018,Garimella_2018,cinelli2021echo}, may form the ideal incubator for the growing of such conspiracy communities \cite{klein2019pathways,phadke2021characterizing}.
The social media platform Reddit, in particular, has been criticized for hosting extreme right-wing content~\cite{gaudette2021upvoting}, and possibly boosting violent and dangerous conspiracy theories from fringe websites into mainstream discourse~\cite{zannettou2017web,rollo2022communities}.

To contrast these harmful dynamics, as well as the diffusion of inappropriate content in general, most social media platforms implemented moderation policies,  i.e. governance mechanisms that structure participation in a community to facilitate cooperation and prevent abuse~\cite{grimmelmann2015virtues}.
These policies {differ in severity, being } based on strategies such as post-warning labels, quarantines, shadow bans, the removal of posts, up to the permanent ban of single users or whole groups responsible for violating the platform's usage policy.
Warning labels are a nudging-like strategy used on several social media to provide platform-mediated information to users about posts~\cite{nassetta2020state, zannettou2021won,ling2022learn}, which, however, may trigger a higher level of engagement~\cite{zannettou2021won,sanderson2021twitter}.
Also quarantining, aimed at preventing direct access to and promotion of controversial communities, has been found to be essentially ineffective in terms of reduction of antisocial behaviors \cite{chandrasekharan2022quarantined}.
\emph{Deplatforming}, instead, is the attempt to limit the danger posed by an individual or a group by removing the platforms (e.g. specific channels, media, or websites) that they use to propagate their content \cite{rogers2020deplatforming}.
On Reddit, for instance, in 2015 five subreddits were closed due to violations of the anti-harassment policy, including \R{fatpeoplehate}---a large subreddit dedicated to weight-based bullying. %

The efficacy of banning groups of users has been evaluated along two different, interdependent directions. 
On the one hand, community-level moderation can be effective \emph{within} the affected platform.
For instance, after the 2015 subreddit ban, users remaining on Reddit significantly reduced their level of hate speech~\cite{10.1145/3134666}.
Also banning influent users who spread conspiracy and hate on Twitter has been shown to reduce their supporters' activity and toxicity \cite{jhaver2021evaluating}.
On the other hand, users of banned communities can also decide to collectively migrate to alternative platforms: the efficacy of content moderation policies should thus be evaluated also by considering effects \emph{between} platforms.
This case was studied in terms of migration to websites, finding an overall reduction of activity despite increased toxicity \cite{10.1145/3476057}, supporting the hypothesis that community bans can also be effective with respect to the broad web ecosystem.
After Reddit’s 2015 ban, affected users massively joined Voat---a news aggregator website explicitly indicated as a safe harbor for communities banned from Reddit \cite{mekacher2022can}. 
Although the effects of community bans have been studied separately within or between platforms, a direct comparison between the behavior of users who migrate to the new platform and those that remain in the old one is still missing. 
{Furthermore, since users who participate in conspiracy communities exhibit a higher engagement compared to other users~\cite{samory2018government, engel2022characterizing}, such a comparison is crucial to understand the impact of deplatforming on this particular group of users.}

To address the lack of research in this area, we quantitatively compare two groups of users affected by a community ban: users who migrated to a new platform and those who remained on the old platform. 
We do so for two communities, the QAnon-focused subreddit \GA, banned in 2018, and the hate-speech-oriented subreddit \FPH, removed in 2015.
Due to the lack of moderation, both communities identified Voat as their main resettlement platform.
Our aim is to determine the impact of belonging to a conspiratorial community on user behavior with respect to deplatforming.

First, we estimate how many Reddit users decide to migrate to Voat and/or completely leave Reddit, following the ban.
We show that the fraction of migrating users who also abandoned Reddit is much higher within the conspiracy-related community.
Next, we analyze the behavior of migrating users by differentiating between those who stayed on the old platform and those who completely left it.
Users who completely left Reddit and were part of a conspiracy community exhibit higher levels of activity and toxicity on Voat compared to the non-conspiracy community. 
A small group of dedicated, migrating individuals played a significant role in the growth of the new \GA community on Voat.
Finally, we observe that the new conspirational community is also able to partially reproduce the old social network: users continue to interact with the same peers after migration, 
{providing evidence for} a more resilient community structure.
Our findings show that deplatforming can effectively reduce the activity, size, and connections with other groups of both types of communities. 
However, conspiracy communities tend to be more resilient, an element which should be taken into account when designing content moderation policies.

\section{Results}

We consider two main data sources, Reddit and Voat, and in particular the communities of \GA and \FPH, described in Section Data.
First, we estimate how many users join the new platform (Voat) and leave the old one (Reddit).
Next, we compare the behavior of users between and within platforms, in terms of joined topical communities, activity, and toxic comments.
Finally, we examine the resilience of the network structure of the two communities.

\subsection{Quantifying migration after deplatforming}
\label{sec:classes}

 Members of a banned community on a social media platform face two binary choices: (i) staying on the old platform or leaving, and (ii) joining a new one to rebuild the banned community or not.
The combinations of these choices define four possible behavioral classes: users (i)~remaining on Reddit \emph{and} joining Voat, (ii)~remaining on Reddit \emph{and} not joining Voat, (iii)~leaving Reddit \emph{and} joining Voat, and (iv)~leaving Reddit \emph{and} not joining Voat. Quantifying migrating users can be challenging and it is often neglected. 
However, estimating the number of users for each behavioral class is crucial to understand the potential interest in a new environment and evaluating the impact of a ban on the broader web.

\begin{figure}[tbp]
\centering
\includegraphics[width=\columnwidth]{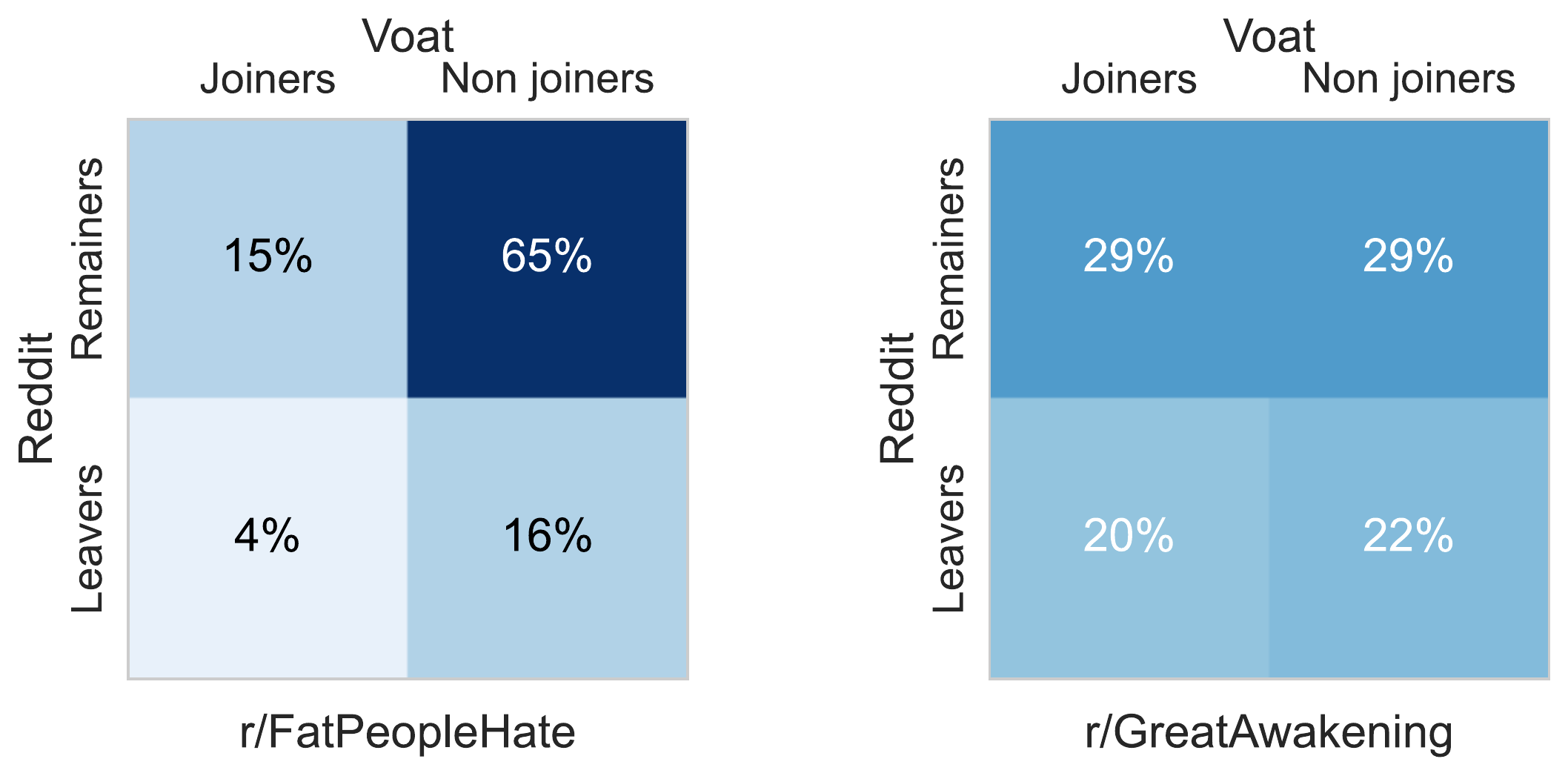}
\caption{Estimated fraction of users in the four behavioral classes: %
Reddit remainers or leavers, and Voat joiners or non-joiners.}
\label{fig:classes}
\end{figure}

To this aim, we combine some measurements performed on the Reddit and Voat data sets with a minimal set of simplifying assumptions, which we describe in detail in the Methods section and briefly recap here.
First, we identify the total number of users participating in the banned subreddits at the moment of the
ban and quantify the fraction of these users who leave Reddit altogether after the ban.
Then, we estimate the number of users joining Voat that might be users of the banned subreddits.
Our estimation is based on two simplifying assumptions. 
First, we assume that new users joining a Voat community in the months following the ban of the corresponding subreddit (in excess with respect to the previous joining trend) are likely Reddit users fleeing the banned community. 
Second, we assume that users with the same username on Reddit and Voat correspond to
the same individuals, and that the behavior of this subset is representative of the general population of their respective community. 
Note that we do not assume that all users would adopt the same username on both platforms, just that a certain fraction of them does it. 
Under this assumption, two identical usernames correspond to the same individual, identifying a subset of the migrating users. 
Note that this assumption has been employed in other studies~\cite{ali2021understanding,10.1145/3476057,newell2016user}, while more sophisticated user matching procedures \cite{b810ebcd29e64cc3959a7363bf78ca84} cannot be performed here.
This assumption is also validated by our further analysis of their network, described later.
These two assumptions allow us to estimate (see Methods) the number of users for each behavioral class, reported as a fraction of the total number of users of \GA and \FPH ($N_{\sGA}= $ \num{24569} and $N_{\sFPH}= $ \num{70739}) in Figure~\ref{fig:classes}.

We observe a considerable difference between GA and FPH in the fraction of users belonging to the four classes. 
First, only 20\% of users (sum of the bottom row) leave Reddit after the FPH ban, while more than 40\% do the same after the GA ban.  
More specifically, only 4\% of users fully migrate from Reddit to Voat after the FPH ban, while 20\% do so for GA.
These users are the most attached to the community banned.
Conversely, the users who continue to use Reddit without joining Voat after the ban, who probably are not much affected by the community ban, are more for FPH than for~GA~(65\%~vs~30\%).

These findings suggest that FPH users may perceive their participation to the community as a casual activity, and not as a part of their identity.
On the contrary, users populating the GA community are much more involved in the subreddit and react very differently to the community ban. 
Overall, the deplatforming seems to affect the behavior of a part of users, but the fraction of the users affected is determined by the importance of the banned subreddit for the users involved.

Of course, the specific numbers we find depend on how ``participation'' is defined: for instance, we can consider users who post at least $n$ messages in the given community.
This way, we would subselect users who are more active, and thus are more likely to appear in the Voat joiners class.
However, we find that the striking difference we observe between the two communities is robust to this change, hinting at a general behavioral difference between the two groups.
The same consideration holds for the choice of the 6 months threshold used to estimate new users joining Voat (see Methods): while changing this threshold changes the number of users in the Voat joiners class, the difference between the conspiratorial and the hate speech communities remains remarkable.

\subsection{Migrating users change their behavior}
\label{sec:behavior}

\begin{figure*}[tbp]
    \centering

\includegraphics[width=0.88\paperwidth]{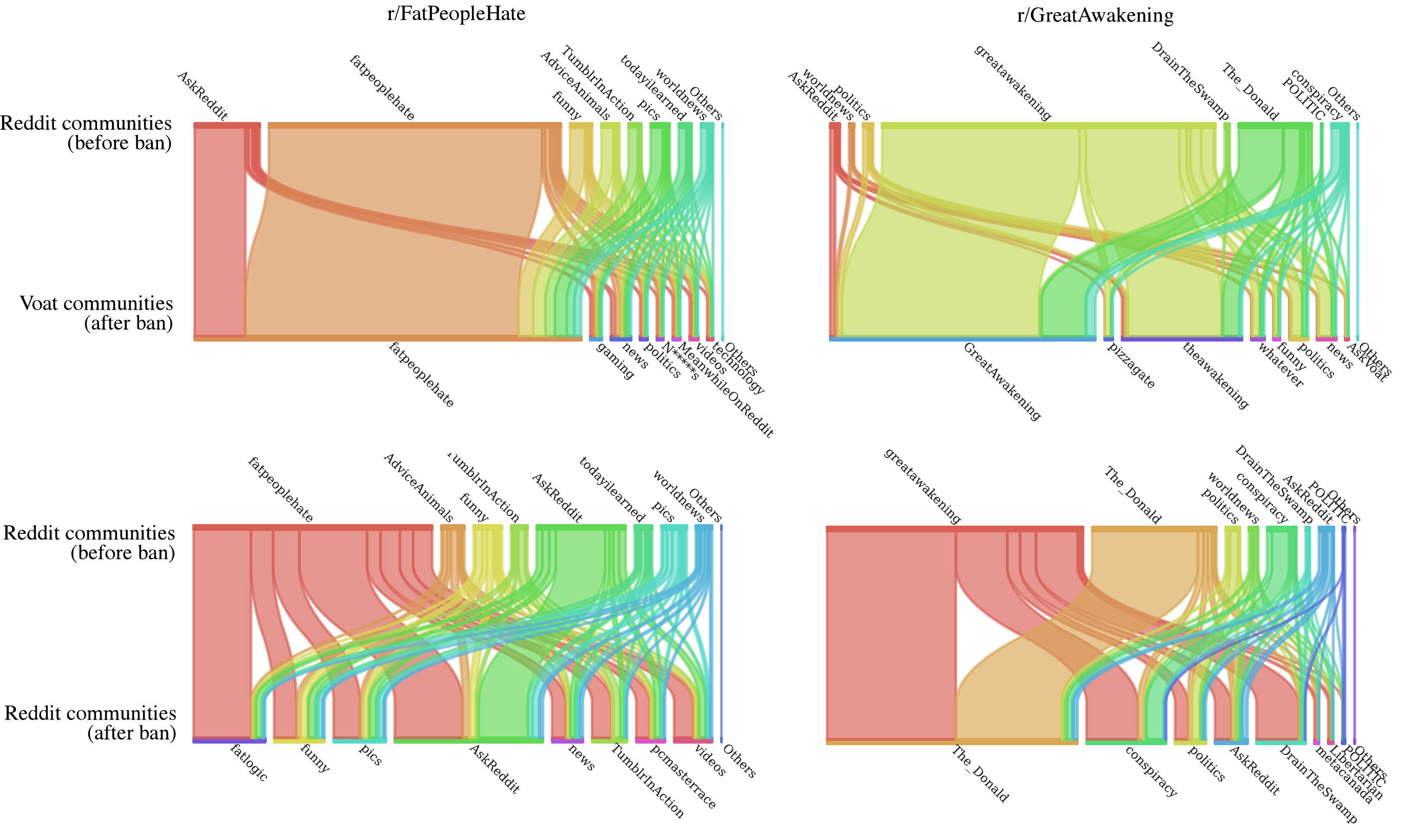}
    \caption{%
    \label{fig:communities}
    The top row shows users migrating from Reddit communities (on the top side of the plot) to Voat communities (bottom side). 
    The bottom row shows users remaining on Reddit, communities joined before the ban (top side of the plot) versus after it (bottom side).
    \FPH is on the left column and \GA on the right.
    In each plot, the width  of each community is proportional to the number of users participating in that community, weighted by the number of messages posted there, in the 6 months before or after the ban. 
    }
\end{figure*}

\begin{figure}[tbp]
    \centering
    \includegraphics[width=\columnwidth]{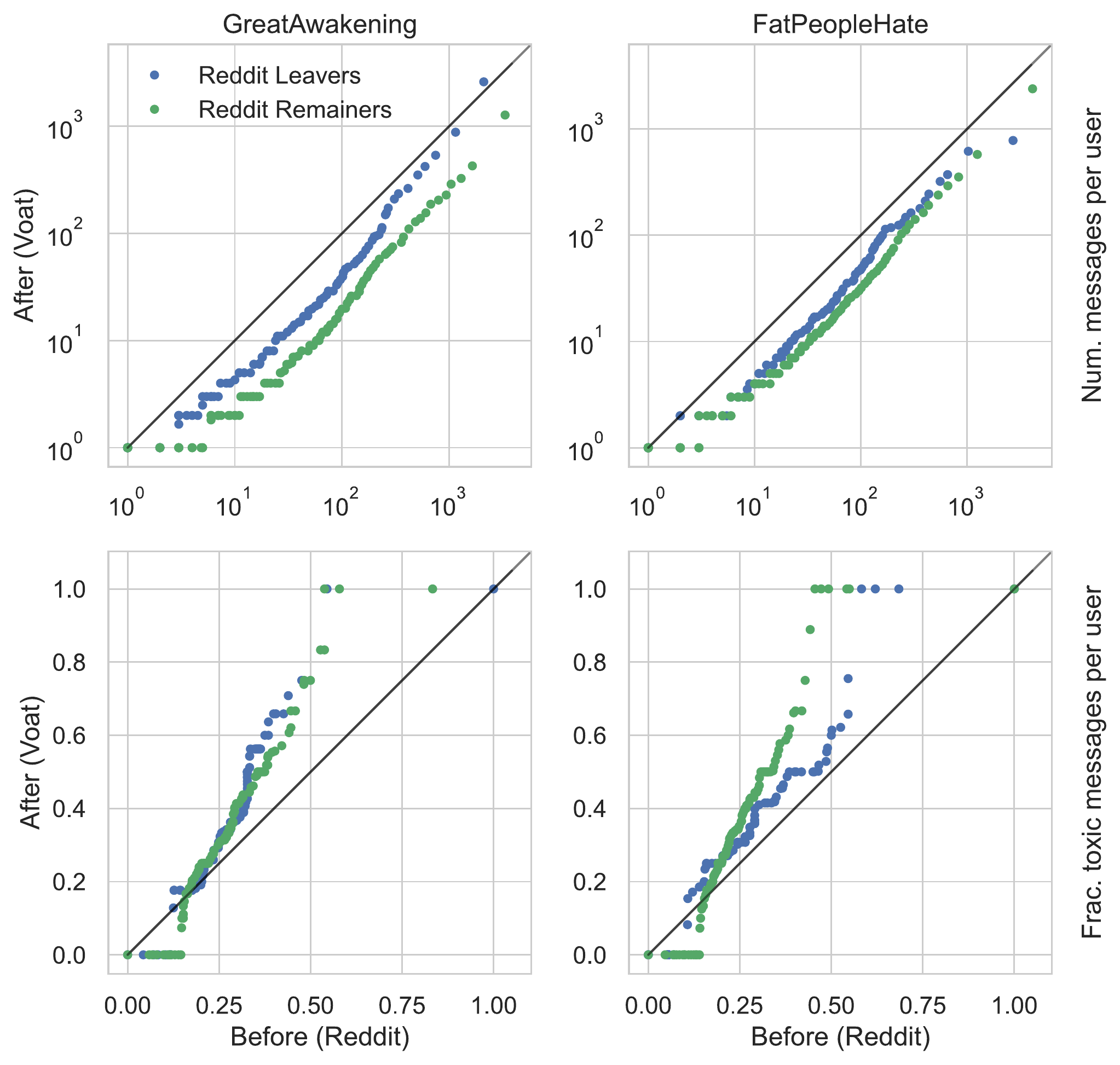}
    \caption{QQ Plot comparing the distribution of the number of messages (top row) and fraction of toxic messages (bottom row) per user on GA (left column) and FPH (right column), in the 6 months before the ban on Reddit (x-axis) and 6 months after the ban on VOAT (y-axis).
    We distinguish between users leaving (blue dots) and remaining (green dots) on Reddit.}
    \label{fig:qqplots}
\end{figure}

We now focus on the behavior of the users joining Voat. 
Since we want to compare their behavior on Voat and Reddit, we only consider users with the same username on both platforms, which we assume to belong to the same individual, as described in the previous section. 
As detailed in Methods, this focus group is composed of \num{1315} users for \GA and \num{2596} for \FPH who use the same pseudonymous on both platforms.
For these users, we are able to gather their entire posting history on Reddit and on Voat.

In this section, we show how the interest in different communities by this subgroup of users changes \emph{within} and \emph{between} platforms (Reddit and Voat).
Then, we investigate whether users leaving Reddit altogether show any difference in behavior with respect to users who keep using both platforms.
To do so, we compare their behavior on Reddit and Voat in the six months before or after the ban, respectively, in terms of participation in other communities, activity, and usage of toxic language.

\spara{Community participation.}
We start by studying how community participation changes for users migrating from Reddit to Voat.
Figure~\ref{fig:communities} (top row) shows the participation of users in the new communities on Voat with respect to the old ones joined on Reddit, for \FPH (left) and \GA (right).
In these Sankey plots, the width of each Voat or Reddit community (on the top and bottom of the plot, respectively) indicates the number of users participating in that community in the 6 months before the ban (for Reddit communities) or after it (for Voat). 
All users have the same weight (set equal to 1), split proportionally to the number of messages posted in the communities they participate in.

We observe that users tend to mainly join new communities on Voat corresponding to the old, banned ones.
For instance, \sFPH users on Reddit mainly join the new \sFPH community on Voat, and the same is true for \sGA users.
However, while \sFPH users join also generalist communities on Voat, such as gaming and technology, \sGA users tend to choose communities related to conspiracy.
For example, 28\% of \sGA users post on \V{Conspiracy}; 11\% on \V{pizzagate} on Voat.
Many \GA users also participate in \V{The\_Donald} (44\%), the community for supporters of Donald Trump.

Then, we perform the same analysis \emph{within} Reddit.
Figure~\ref{fig:communities} (bottom)  shows how participation to communities on Reddit shifts to other Reddit communities after the ban, for users remaining in Reddit.
Users in the conspiracy community massively move to \R{The\_Donald} (70\% of users do so) and \R{conspiracy} (38\%), as well as \R{DrainTheSwamp} (29\%), another trumpist and conservative subreddit~\cite{massachs2020roots}.
On the contrary, only 6\% of users from \sFPH participate in a community with a topic similar to the former one (\R{fatlogic}) after the ban.

\spara{Activity.}
Then, we compare the number of messages posted on Reddit (pre-ban) and on Voat (post-ban).
Within this comparison, we differentiate between two groups of migrating users: those who keep participating on both platforms and those who leave Reddit.
Figure \ref{fig:qqplots} (top row) shows results for \sGA and \sFPH as a quantile-quantile (QQ) comparison: %
if two distributions are similar the quantiles will lie on the diagonal; dots below the diagonal indicate a decrease of the quantity on the y-axis with respect to that on the x-axis, dots above indicate an increase.
Figure \ref{fig:qqplots} (top row) shows that the activity on the new platform is generally lower than on the old one, for both very active and less active users.
This observation holds for both communities, \GA (left column) and \FPH (right column).
One possible explanation is that since the new-borne community is still in its infancy, it generates less engagement.
This observation suggests that, to some extent, deplatforming is effective in reducing engagement in both communities.

However, the figure shows that for \GA Reddit leavers are more active on the new Voat community than users remaining on Reddit.
In fact, Reddit remainers post on average $7.0$ messages per month, compared to $11.3$ for leavers.
Users who fully migrate to the new conspiracy community are therefore relatively more engaged with the new platform.
In particular, very active \sGA users (top $1\%$) leaving Reddit are even more active on the new platform, despite the average activity on Voat being much lower  (see the blue dot above the diagonal in the left panel of Figure \ref{fig:qqplots}).
In other words, the growth of the new \sGA community on Voat is mainly driven by a few ``zealots'' migrating to the new platform, while this effect is absent on \sFPH, a non-conspiracy community, where most active users are less active on Voat (the blue dots below the diagonal in the right panel of Figure \ref{fig:qqplots}).

The difference between leavers and remainers is in fact less prominent for \FPH: the average number of messages per month is $7.3$ for remainers and $8.3$ for leavers.
In order to assess whether remainers and leavers are indeed significantly different in terms of change in activity, we perform a two-sided Mann-Whitney U statistical test, to compare the ratio of messages between the new and the old platform for leavers and remainers.
This test shows that the two groups are very different for \sGA ($U$ statistics of $5.2 \cdot 10^4$, $p$-value $< 10^{-4}$).
For \sFPH, the two groups are also different but much less visibly ($U$ statistics of $1.9 \cdot 10^5$, $p$-value $0.03$).

\spara{Toxicity.}
Next, we compare the use of toxic language between platforms, again distinguishing Reddit leavers and remainers.
Quantifying toxic messages is relevant to assess whether a less regulated environment induces an increase in the use of toxic speech.
To this aim, we classify each message with the IMSyPP classifier~\cite{kralj2022handling} (see Methods).
Figure \ref{fig:qqplots} (bottom row) shows a QQ comparison of the fraction of toxic messages between platforms.
We observe that the level of toxicity is generally much higher on the new Voat communities than on the corresponding banned subreddits.
In particular, the average fraction of toxic comments increases by 8 percentage points for \sGA (from $28\%$ to $36\%$) and by 10 points for \sFPH (from $25\%$ to $35\%$).
The figure shows a more  pronounced increase for users with the largest fraction of toxic messages, on both \sGA and \sFPH communities.
Furthermore, the average increase of toxicity is different between leavers and remainers, but the difference is reversed in the two communities {(higher for leavers on GA and for remainers in FPH)}.
For \sGA, such an increase is $+10.3$ points for leavers and $+7.5$ points for remainers.
For \sFPH, it is $+5.2$ for leavers and $+10.5$ for remainers.
In both cases, the difference between the two groups is significant according to a two-sided Mann-Whitney U statistical test ($p < 10^{-5}$).

Therefore, it appears that deplatforming a conspiracy community strongly increases the toxicity of users completely leaving the old platform; while this effect is less evident for a non-conspiracy community.
We speculate that the observed difference might be due to a form of polarization, possibly strengthened by 
a narrative of events shared by the conspiracy community, which eventually results in a permanent departure from the old platform. 
Contrarily, for a hate speech community, the toxic behavior may be unrelated to polarization but simply to the removal of constraints in the less-moderated platform.

\begin{figure}[tbp]
    \centering
    \includegraphics[width=0.5\textwidth]{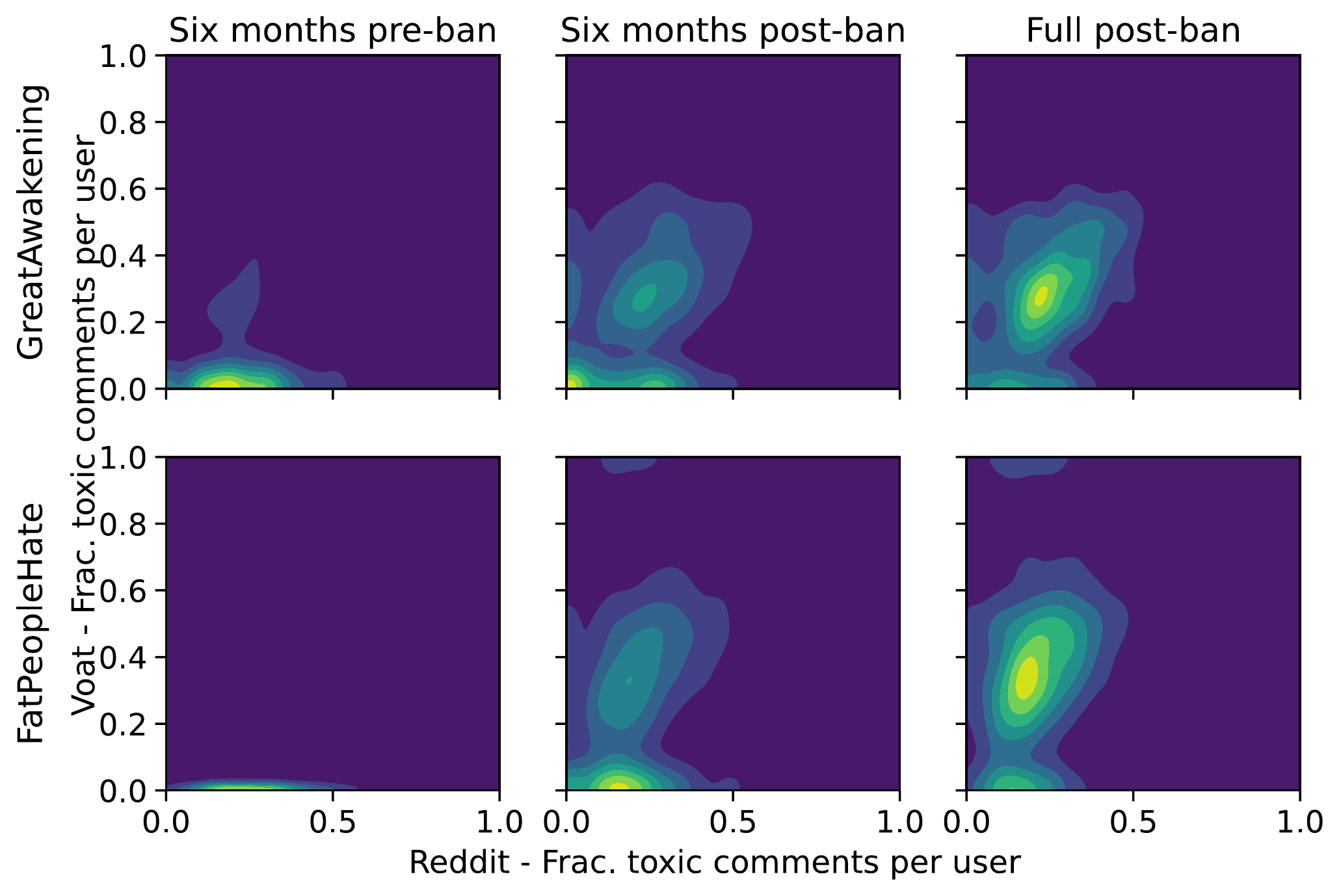}
    \caption{Temporal evolution of the levels of toxic speech on Reddit (x-axis) vs Voat (y-axis), on \GA (top row) and \FPH (bottom row). We distinguish between the fraction of toxic messages on Voat six months before the ban of the subreddit (left column), six months after it (middle), and up to two years after it (right column).}
    \label{fig:hs_evo}
\end{figure}

We investigate this phenomenon in more detail by disentangling the temporal evolution of the toxic speech levels for users belonging to GA and FPH. 
Figure~\ref{fig:hs_evo} shows the density of toxic messages in the Reddit vs Voat phase diagram, for GA (top row) and FPH (bottom row). 
We distinguish between messages on Voat six months before the ban date (left column), six months after the ban (middle), and up to two years after the ban (right column). 
This figure shows that as soon as the new platform gets populated, the number of toxic comments on the new Voat community increases with respect to the banned subreddit. 
In more detail, the majority of users posted about $20\%$ of toxic messages on \GA on Reddit (slightly less on \FPH), while they post almost $40\%$ of toxic messages on Voat.

\subsection{Measuring community resilience to bans}
\label{sec:network}

\begin{figure}[tbp]
    \centering
    \includegraphics[width=\columnwidth]{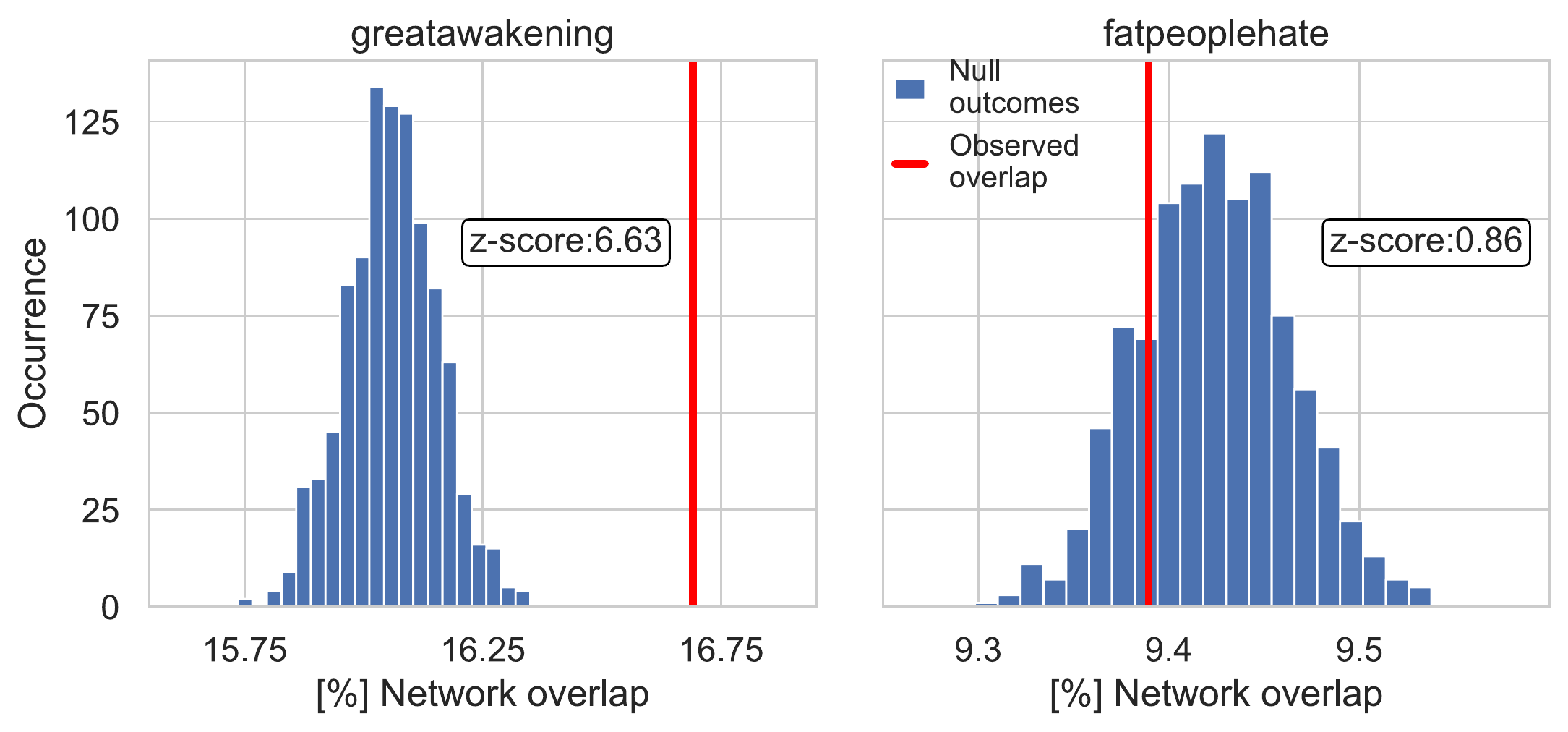}
    \caption{Network overlap in the \R{greatawakening}~(left) and \R{fatpeoplehate}~(right) communities~(red bar), compared with a null model reshuffling the interactions~(blue histogram). 
    In \GA, the observed overlap is not compatible with random fluctuations. %
    Conversely, \FPH shows an overlap fully compatible with random re-organization.}
    \label{fig:gra_fph_netw_overlap}
\end{figure}

The results displayed so far suggest a difference between the users participating to FPH and GA in terms of their response to the community ban that could be reconducted to a different sense of community belonging.
We further test this observation by means of a network-based analysis, i.e. we compare the two social networks on the banned subreddits and on the corresponding communities on Voat, where two users are connected if they commented on the same post (see Methods).
As we are considering users that appear on both social media platforms, we can investigate the overlap in the structure of the two networks, that is the social relations of the banned community that are preserved after the migration to Voat.
We find that 16.7\% of the GA banned community is reconstructed on Voat, while this fraction is 9.4\% for FPH.
We test the significance of these values by building a null model which shuffles the set of posts commented by each user, see Methods.
Figure~\ref{fig:gra_fph_netw_overlap} shows that the empirical overlap for GA is not compatible with the null model (z-score: 6.6), while it is for FPH (z-score: 0.86).

First, this finding consolidates our assumption that users with the same username on both platforms correspond to the same individual, since these results cannot be attributed to chance.
Then, this finding confirms that the conspiracy community is more resilient to deplatforming:
the social engagement created by sharing the same narrative helps conspiracy users to reconstruct even the user-user interactions of the banned community. 
A possible mechanism that could explain this effect is that the broad narrative behind the GA conspiracy theory is tied to the study and discussion of a number of sub-issues and topics.
Indeed, sub-communities of users might gather around such topics, leading to a resilient community structure in \GA.
Conversely, in \FPH,
where no shared knowledge has to be preserved or reinforced, the observed overlap is fully compatible with random fluctuations.

\section{Discussion}

In this paper, we investigated the difference in the behavior of participants of deplatformed subreddits, comparing a conspiracy and a non-conspiracy community. 
{The conspiracy narrative is one of the most recognizable aspects characterizing online discussions~\cite{samory2018government} and --- despite the number of conspiracy posts being negligible with respect to other categories, (e.g. news~\cite{del2016spreading,vosoughi2018spread}) --- conspiracy still finds room for active online debates.}
We find that users in the conspiracy community are much more likely to leave the moderated platform altogether and join a new, unmoderated one. %
Furthermore, while both communities display a decrease in activity, the decrease is less prominent for the conspiracy community.
Users migrating from the deplatformed conspiracy community are more likely to recreate their social ties on the new platform.

Overall, our findings point to a higher resilience --- i.e., the ability to recreate the old community after a disruption --- of conspiracy groups, that may be due to a different involvement with the narratives that define the two groups.
Since conspiracy narratives strongly affect the sense of self of their participants~\cite{hacker2021socio}, they are likely to develop a strong attachment to the community, to its norms, narratives and values, with respect to users casually participating in a hate speech community. 
Moreover, an effort in preserving this set of common values and knowledge might explain their ability in reconstructing social ties.
{Our findings highlight new aspects of participation in conspiracy communities, in line with other results regarding conspiracy theories on Reddit about the mechanisms that drive users to join~\cite{klein2019pathways}, participate~\cite{phadke2022pathways} and leave conspiracy communities~\cite{phadke2021characterizing}.
In a nutshell, users that engage with conspiracy theories display a set of detectable features including linguistic and social ones. }

More work is needed to carefully evaluate these aspects highlighted by our work, and to overcome its limitations. 
For instance, our method for finding common users across platforms --- i.e., simply matching their usernames --- could be improved by adopting more complex techniques, such as text mining and author detection.
Furthermore, our analysis considers only two communities: while data availability constrained this choice, other comparisons of conspiracy vs non-conspiracy communities may be undertaken in the future.
Similarly, migration to other platforms besides Voat could be taken into account, by following our framework.
Moreover, our study is purely observational: a causal analysis of the relationship between user behavior, participation in conspiracy communities, and deplatforming would be needed.
However, our work is a necessary first step, as it shows that such a link could provide valuable insights into the effect of moderation policies on conspiracy communities.
Finally, a more in-depth investigation of the preservation of norms, values, and narrations during the migration could further corroborate our hypothesis.

Our findings have implications combining both technical and social aspects. %
Social media platforms must weigh in the effectiveness of severe moderation with the potential social costs it brings, such as migration to alternative platforms and heightened user toxicity. 
With indications that conspiracy thinking is rising, possibly also due to growing economic disparities~\cite{jetten2022economic}, social media platforms should thoroughly evaluate counter-measures.

\section{Data and methods}

In this section, we describe in detail the data set used and the methodology to estimate the number of users migrating from Reddit to Voat.

\subsection{Data}
\label{sec:data}

Here we provide an overview of our two main data sources: Reddit and Voat, and in particular the two communities we analyze, \GA and \FPH.

\spara{Reddit.}
Reddit is a social content aggregation website, organized in topical communities, called \textit{subreddits}, centered around a variety of topics where all users must have a pseudonymous account in order to participate.
Users can post submissions in these subreddits, and comment on other submissions and comments, thus creating a tree structure for the overall discussion.
We call a message a generic piece of user-generated content, when the distinction between submission and comment is not relevant.
In addition, users can also upvote a submission to show approval, appreciation, or agreement (and their opposites with a downvote).
The score of a submission is the number of positive votes minus the number of negative votes it has received.
Differently from other social media like Facebook or Twitter, Reddit's homepage is organized around subreddits, and not on user-to-user relationships.
As such, the subreddits chosen by a user represent the main source of the information they consume on the website.

Reddit has already been studied within a rich set of research frameworks. 
For instance, the study of user engagement and interactions between highly related communities~\cite{tan2015all,hessel2016science}, the analysis of post-election political analyses \cite{barthel_2016}, or the impact of linguistic differences in news titles \cite{horne2017impact}.
Interestingly, it has already been used to explore harmful social dynamics on online social media such as hate speech \cite{saleem2017web} or cyberbullying \cite{rakib2018using}.
Health-related issues have also been studied on Reddit, like mental illness \cite{de2014mental}, as well as the opioid epidemics in the U.S.~\cite{park2017towards,balsamo2019firsthand}.
We collect public data from Reddit using the Pushift collection~\citep{baumgartner2020pushshift}.

\spara{Voat.}
Voat.co was a news aggregator website, shut down on December 25, 2020, that has been indicated as a safe harbor for communities banned from Reddit~\cite{mekacher2022can}.
Like Reddit, discussions on Voat occurred in specific groups of interests  called  “subverses.”
Voat has been also defined as a ``clone'' of Reddit,\footnote{\href{https://www.theguardian.com/technology/2015/jun/22/reddit-clone-voat-dropped-by-web-host}{The Guardian, \emph{``Reddit clone Voat dropped by web host for 'politically incorrect' content''}.}} since it mimicked its functionalities and interface:
users  can  subscribe  to subverses of interest, comment, upvote, and downvote the comments  and  submissions in a tree-like commenting system.
In this paper, we use the dataset collected by \citet{mekacher2022can}. 

\spara{\GA.}
Among the banned communities on Reddit that reportedly migrated on Voat, an example of particular interest is \rGA, a subreddit dedicated to the diffusion of the QAnon conspiracy theory.
This subreddit has been identified as the largest QAnon-related community on Reddit~\cite{papasavva2022gospel}:
it had an active userbase with over 71\,000 subscribers and an average of 10\,000 comments per day, and it was banned on the 18th September 2018 for repeated content violations.
In particular, the community was responsible for harassing a user that was misidentified as the suspect of the Jacksonville Landing shooting.
On Voat, \vGA attracted a large interest, being one of the most popular subverses with a very engaged community \cite{papasavva2021qoincidence}.

\begin{figure*}
    \centering
        \includegraphics[width=0.43\textwidth]{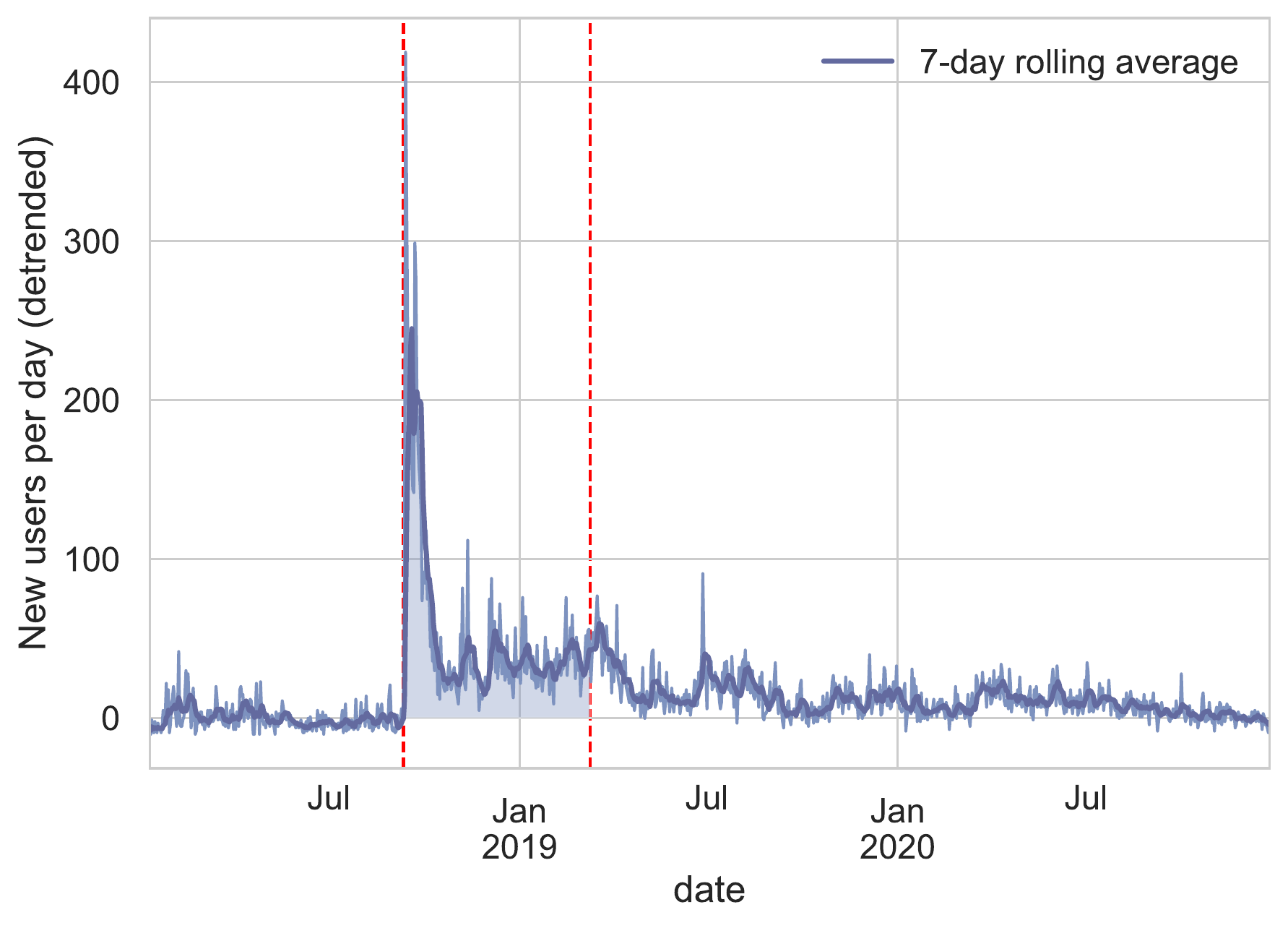}
        \label{fig:new-voat-users-ga}
    \hfill
        \includegraphics[width=0.44\textwidth]{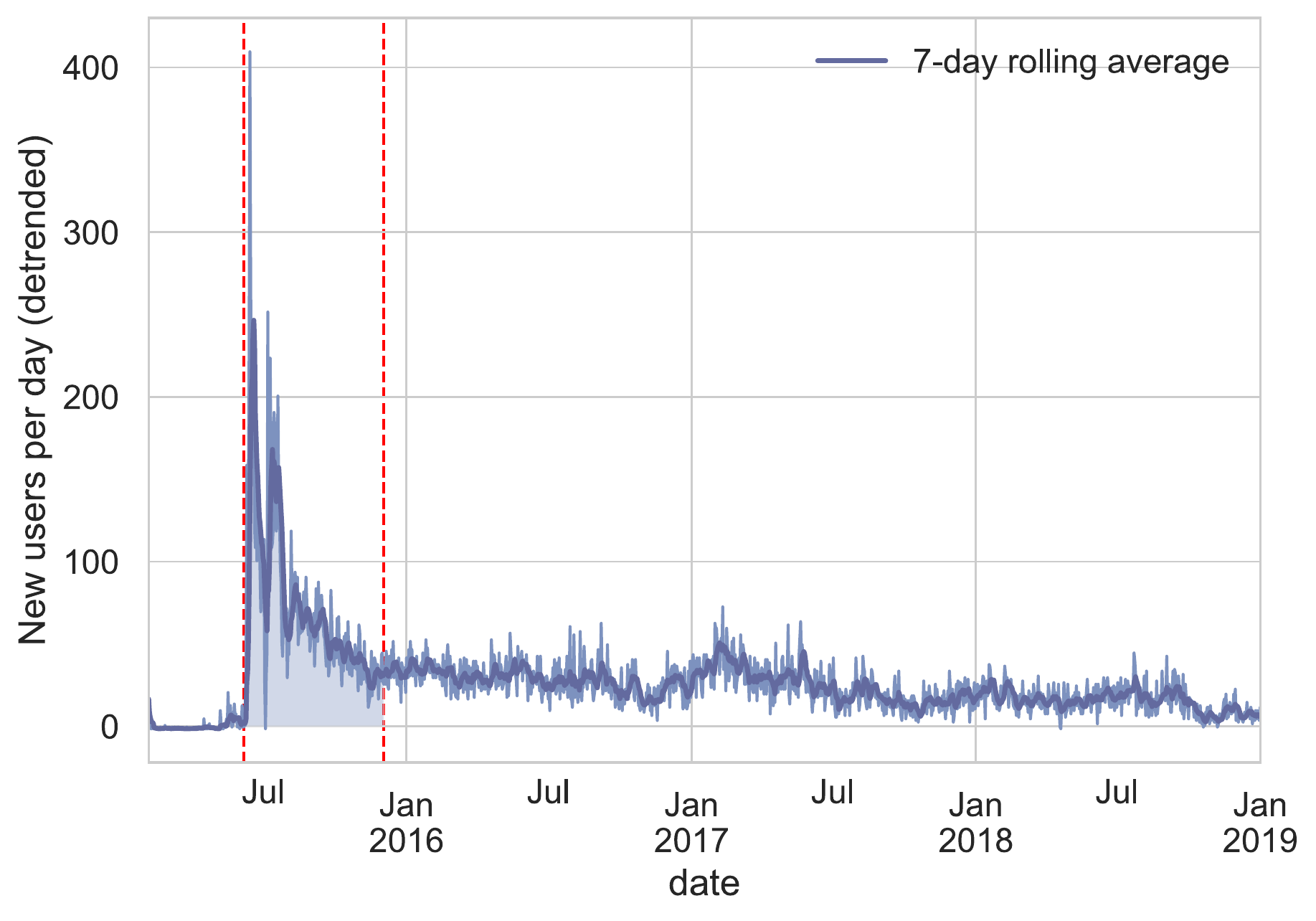}
        \label{fig:new-voat-users-fph}
    \vspace{-4mm}
    \caption{Number of new users per day in the Voat communities of \GA (left) and \FPH (right) after detrending (i.e., removing the average pre-ban number of new users per day). In each plot, the first red tick indicates the date of the ban of the corresponding Reddit community and the second red tick corresponds to 6 months after such ban date.}
    \label{fig:new-voat-users}
\end{figure*}

\spara{\FPH.}
In order to study the difference in the effects of deplatforming between a conspiracy-related group, \GA, and other types of communities, we consider \rFPH, a subreddit banned for hate speech, as a comparison.
While many other communities have been banned over the years, this subreddit is the only one that is, at the same time, non-related to conspiracies, and for which anonymized data is available for both Reddit and Voat.
\rFPH was a subreddit entirely dedicated to body-shaming individuals, by posting pictures of overweight people that were then ridiculed.\footnote{\href{http://www.theverge.com/2015/6/10/8761763/reddit-harassment-ban-fat-people-hate-subreddit}{The Verge, \emph{``Reddit bans 'Fat People Hate' and other subreddits under new harassment rules''}.}}
The subreddit internal rules prohibited users from expressing ``fat sympathy''.\footnote{\href{https://www.washingtonpost.com/news/the-intersect/wp/2015/06/10/these-are-the-5-subreddits-reddit-banned-under-its-game-changing-anti-harassment-policy-and-why-it-banned-them/}{Washington Post, \emph{``These are the 5 subreddits Reddit banned under its game-changing anti-harassment policy — and why it banned them''}.}}
On 6th June 2015, the subreddit was banned: with over 150\,000 subscribers at the time of the ban, it has been one of the largest banned subreddits.
As in the case of \GA, previous work reported that part of \FPH users migrated to Voat~\cite{mekacher2022can}, with the growth of an analogous subvoat with the same name after the Reddit ban.

\subsection{Methods}

\spara{Estimate the number of migrating users.}
The ban of a community by a social media platform can be answered by its members with two binary choices: they can decide to (i)~remain or leave the old platform, and (ii)~join or not a new one to reconstruct the community banned.
We indicate by $\mathcal{U}_s$ the set of all users  participating in the banned subreddit $s$ at the moment of the ban, with cardinality $N_s$.
We indicate by $\mathcal{R}_s$ the subset of $\mathcal{U}_s$ formed by users who keep participating in Reddit after the ban date, with cardinality $R_s$.
Conversely, users who leave Reddit after the ban are the complement set of $\mathcal{R}_s$, i.e. 
$\overline{\mathcal{R}}_s = \mathcal{U}_s \setminus \mathcal{R}_s$, with cardinality  $|\overline{\mathcal{R}}_s| = N_s-R_s$.
In the same way, we indicate by $\mathcal{V}_s$ the subset of $\mathcal{U}_s$ formed by users who join Voat after the ban date ($|\mathcal{V}_s| = V_s$), and by $\overline{\mathcal{V}}_s = \mathcal{U}_s \setminus \mathcal{V}_s$ the complement set of users not joining Voat ($|\overline{\mathcal{V}}_s| = N_s-V_s$). 

The four possible behavioral classes are defined as the intersection of these sets: users (i)~remaining on Reddit \emph{and} joining Voat, $\mathcal{R} \cap \mathcal{V}$, (ii)~remaining on Reddit \emph{and} not joining Voat, $\mathcal{R} \cap \overline{\mathcal{V}}$, (iii)~leaving Reddit \emph{and} joining Voat, $\overline{\mathcal{R}} \cap \mathcal{V}$, and (iv)~leaving Reddit \emph{and} not joining Voat, $\overline{\mathcal{R}} \cap \overline{\mathcal{V}}$.
Our first goal is to provide a tentative estimate for the number of users for each of these four behavioral classes.
Indeed, quantifying migrating users is a generally overlooked task, that can provide us estimates of the potential interest towards a new environment and, in turn, the effectiveness of the ban with respect to the broad web environment.
To this aim, we combine some measurements performed on the Reddit and Voat data sets with a minimal set of simplifying assumptions.

First, we identify the total number of users participating in the banned subreddit at the moment of the ban by counting users posting at least one submission or comment in the previous 6 months.
This count gives us $N_{\sGA}= $ \num{24569} user for \GA and $N_{\sFPH}= $ \num{70739} users for \FPH.
Then, we detect how many users in this set are still active on Reddit in the 6~months following the ban date of each community, over the whole Reddit, to quantify the users leaving Reddit altogether. 
This gives us the number of Reddit remainers, i.e. users that show non-zero activity (as submission or comment) in any subreddit after the ban.
We obtain $R_{\sGA}= $ \num{14243} for \GA and $R_{\sFPH}= $ \num{56918} for \FPH.

Next, we wish to estimate the number of users joining Voat that might be users of the banned subreddits. 
To do so, first, 
we count the number of daily new users in the $\sGA$ and $\sFPH$ communities on Voat; then, we detrend this quantity by subtracting the average pre-ban number of daily new users.
We report this time series in Figure~\ref{fig:new-voat-users}.
This plot shows a large spike around the respective Reddit ban date of each community, suggesting a prominent effect of the Reddit ban on Voat.
After the spike, the influx of new users rapidly decreases but remains higher with respect to the baseline for some time. 
Consequently, one can assume that users joining Voat mostly come from banned subreddits, at least for some time after the ban date. 
While it is not possible to precisely estimate when the effect of the Reddit ban ends, one can see that the number of daily new users on $\sFPH$ became steady 6 months after the Reddit ban (Figure ~\ref{fig:new-voat-users}b).
By following these observations, we assume that 
the number of (excess) new users joining a Voat community in the months following the ban of the corresponding subreddit are most likely Reddit users fleeing the banned community.
This is our first simplifying assumption.
To meaningfully compare the two communities, we choose six months as the defining interval for both groups.
We discuss the robustness of this assumption below.
Therefore, we interpret the shaded area in Figure~\ref{fig:new-voat-users} as the number of Reddit users migrating on Voat, obtaining $V_{\sGA}= $~\num{12049} and $V_{\sFPH}= $~\num{13342}.

At this point, for each subreddit $s$, we know how many users remain on Reddit ($|{\mathcal{R}}_s| = R_s$), or leave it altogether ($|\overline{\mathcal{R}}_s| = N_s - R_s$), join Voat ($|{\mathcal{V}_s}| = V_s$) or not ($|\overline{\mathcal{V}}_s| = N_s - V_s$).  
The only missing information to estimate the size of the four behavioral classes is how many users remain on Reddit \emph{and} join Voat, i.e. the cardinality of the intersection $\mathcal{R} \cap \mathcal{V}$. 
The other classes' sizes can be easily derived by subtraction, as detailed later.

To this aim, we consider a subset of the migrating users, namely users who appear with the same username on a subreddit and, after the ban date, on the corresponding community on Voat.
These users are $n_{\operatorname{GA}} = \num{1315}$ and $n_{\operatorname{FPH}} = \num{2596}$.
Our second simplifying assumption is that these users with the same username correspond to the same individuals, and that the behavior of this subset is representative of the general population of their respective community.
We thus calculate the fraction of users in the subsets $n_{\operatorname{GA}}$ and $n_{\operatorname{FPH}}$ who also belong to the set $\mathcal{R}$, i.e. who are still active on Reddit, as users posting on any other subreddit after the ban of their community.
These fractions read $\rho_{\sGA}= 0.59$ and $\rho_{\sFPH}= 0.80$: 
the difference in these numbers is the first suggestion that the two communities show a fundamental difference in the behavioral response to their deplatforming.

By assuming that these fractions are the same for all users remaining on Reddit {and} joining Voat, the number of users in this class is simply $|\mathcal{R}_s \cap \mathcal{V}_s| = \rho_s \cdot V_s$.
Conversely, the number of users abandoning Reddit in favor of Voat will be $|\overline{\mathcal{R}}_s \cap \mathcal{V}_s| = (1 - \rho_s) \cdot V_s$.
The cardinality of the two last sets (users not joining Voat) is obtained by subtracting these numbers from the respective totals.
The number of users remaining on Reddit and not joining Voat is $|\mathcal{R}_s \cap \overline{\mathcal{V}}_s| = R_s - (\rho_s V_s)$, while the number of users leaving Reddit and not joining Voat is $|\overline{\mathcal{R}}_s \cap \overline{\mathcal{V}}_s| = N_s - R_s - (1 - \rho_s) V_s$.

\spara{Toxic comments detection.}
In order to detect toxic comments, we used the IMSyPP classifier~\cite{kralj2022handling} (publicly available~\footnote{https://huggingface.co/IMSyPP}) to label each comment on Reddit and Voat as toxic or not.
We define a message as toxic if it is classified into one of the following categories: (i) \textit{inappropriate}, the message contains terms that are obscene or vulgar, but the text is not directed to any person or group specifically; (ii) \textit{offensive}, the comment includes offensive generalization, contempt, dehumanization, or indirect offensive remarks; or (iii) \textit{violent}, the comment’s author threatens, indulges, desires or calls for physical violence against a target; it also includes calling for, denying or glorifying war crimes and crimes against humanity. If the message is classified in the remaining category, namely \textit{appropriate}, it is considered non-toxic. 
The accuracy for toxicity classification of the IMSyPP classifier is 0.84 \cite{kralj2022handling}. 
As a test of robustness, we contrast the results of this classifier with those of the Google Perspective API, whose toxicity score is in [0,1].
Results are in good agreement, obtaining a ROC AUC between the two classifiers of 0.94.

\spara{Social network analysis.}
We start by reconstructing a bipartite network, linking posts $p$ and users $u$ that commented on them, separately on each platform: we denote these networks as $\mathcal B_{R}(p, u)$ on Reddit, and $\mathcal B_{V}(p, u)$ on Voat.
Next, we consider the unimodal projection on the user side of each bipartite network: we obtain two networks of users where users $u$ and $v$ are connected if and only if they commented on the same post, on the banned subreddit or on the corresponding community on Voat. %
We indicate these two user networks as $\mathcal{G}_R=(\mathcal{N}_R, \mathcal{E_R})$ and $\mathcal{G}_V=(\mathcal{N}_V, \mathcal{E_V})$, respectively.
As we are considering users that appear on both social networks, we can investigate the overlap in the structure of the two networks. 
We define the overlap in the structure of the two networks as the fraction of edges of the Reddit network that are present also in the Voat one, that is $\frac{|\mathcal{E_R} \cap \mathcal{E_V}|}{|\mathcal{E_R}|}$. 
We obtain that 16.7\% of the GA banned community is reconstructed on Voat, while this fraction is 9.4\% for FPH. 

To assess the significance of these values we built a null model that randomizes the Voat bipartite network $\mathcal B_V(p,u)$, by shuffling the set of posts commented by each user thus creating a new bipartite network called $\mathcal B'_V(p,u)$. 
Afterward, we perform the projection of the randomized bipartite network $\mathcal B'_V(p,u)$ and we compute the overlap with the projection obtained by the empirical bipartite network obtained from Reddit $\mathcal B_R(p,u)$. 
Note that the degrees of the nodes are preserved in the randomization.
We repeat the randomization process \num{10000} times and we obtain a Gaussian-like distribution of null overlap values.

\bibliographystyle{unsrtnat}
\bibliography{references}

\end{document}